# ON PERFORMANCE ANALYSIS OF AMBR PROTOCOL IN MOBILE AD HOC NETWORKS

**TANAY DEY, M.M.A. HASHEM AND SUBROTO KUMAR MONDAL**

*Department of Computer Science and Engineering, Khulna University of Engineering and Technology (KUET), Khulna 9203, Bangladesh.*

*E-mails: dtanay2004@yahoo.com, mma_hashem@hotmail.com, subrota_002@yahoo.com*

**ABSTRACT**: Due to mobility of nodes in ad hoc networks, the most challenging issue is to design and to make sound analysis of a routing protocol that determines its robustness to deliver packets in low routing packet overhead. In this paper, we thoroughly analyzed the Adaptive Monitor Based Routing (AMBR) protocol by varying different parameters that affect a routing protocol to measure its performance. Analysis shows that it requires less routing control overhead comparing with other prevalent routing protocols. An improved analytical model is also presented in this paper. All these analyses firmly prove that AMBR is a sound and robust protocol in terms of flooding, routing overhead and hence, enhances reliability.

## 1. INTRODUCTION

An ad hoc network is a class of wireless systems that consists of independent mobile nodes communicating with each other over wireless links, without any static infrastructure such as base stations [1]. A communication session is achieved either through a single-hop radio transmission if the communicating parties are close enough, or through relaying by intermediate nodes otherwise [2]. Since the nodes move randomly, the topology of the network changes with time. Dynamically changing topology and lack of centralized control make the design of an adaptive distributed routing protocol challenging [1]. Due to the limited spectrum, user's mobility and power constraints, routing remains a challenge, particularly in wireless communication systems such as ad hoc networks. Several other challenges complicate routing, including scalability, routing efficiency, adaptation to wireless networks of various densities, and distribution [3].

In this paper, a thorough analysis is performed on the Adaptive Monitor Based Routing (AMBR) [10] protocol to show its robustness by varying different parameters that affect most of the routing protocols. AMBR discovers and maintains routes in hierarchical and distributed fashion and locally repairs the broken link. The main motivation behind the proposed AMBR is to drastically cut down flooding, to substantially tame the routing overhead, to repair broken link locally in order to minimize the routing overhead, and to increase efficiency in packet movement in the ad hoc networks. In all analyzing cases, it is found that AMBR is a bandwidth efficient routing protocol as the routing overhead was drastically cut down.

In section 2, we describe the related work and the details of AMBR routing protocol and its algorithm is described in section 3 and in section 4, we describe an improved analytical model. Simulation results are shown in section 5. Finally, section 6 concludes the paper.

## 2. RELATED WORK

In CBRP [4], problem with having explicit cluster heads is that routing through cluster heads creates traffic bottlenecks. In Landmark, LANMAR [5] and L+ [6], this is partially solved by allowing nearby nodes to route packets instead of the cluster head, if they know the route to the destination. All of the above schemes have explicit cluster heads, and all addresses are therefore relative to these and are likely to have to change if a cluster head moves away [7]. In Janitor Based Routing [12], Janitor works as the cluster head in a cluster but the Janitor selection algorithm is complex and the solution for dynamically route broken is not apt. Geographical forwarding techniques are used for routing in NoGEO [13]. It embeds the network graphs in a virtual two-dimensional co-ordinate space. The main pitfalls of this scheme are that it will only work on certain types of graphs and it has never been evaluate for more than low mobility speed. GEM [14], is another coordinate based routing which embeds a sensor network graph in a polar coordinate system. It confronts a heavy concentration of traffic around the root node as it does tree-based routing. Actually it was designed for sensor networks.

In Zone Routing Protocol (ZRP) and Fisheye State Routing (FSR), nodes are treated differently depending on their distance from destination and it incurs less overhead at the cost of decreased precision [7]. DSDV [8], due to its periodic updates and flat routing tables, experiences very high overhead growth as the networks beyond 100 nodes, but nevertheless performs well in comparison with other protocols in the size ranges studied. AODV [9], due to its reactive nature, suffers from high overhead growth both as the size of the network, and the number of flows, grows. While AODV performs very well in small networks, the trend suggests that it is not recommendable for larger networks [7].

## 3. AMBR ROUTING PROTOCOL

Our proposed protocol AMBR uses nodes which are called monitors, whose first hop connectivity (total no of neighbors) acquire a predefined number over the whole network. If a node gets alive, it broadcasts a message named "hello". This "hello" message is not periodic, rather it is event driven. Any node, getting this message from another node should give back a reply to that node only (not a broadcast). Monitor detection is completely individual responsibility and it is done in every node after gathering the information of the whole network. If a node has calculated that it's total number of neighbors or first hop connectivity is equal to a predefined number then it will broadcast a "new monitor" packet and all its covered nodes will accept it and the "hello" propagation immediately ends as it is no longer an ordinary node, hence it will now act as a monitor. In the case of rapid network change, number of neighbors of a monitor changes consequently but in our technique the monitor does not need to broadcast extra packet to inform its neighbors about this change. In the network, monitor available messages can be piggy backed with any data or acknowledge packet that has gone to monitor from that node and vice versa. Any node that has not sent any data to monitor and has not got any data from monitor over a predefined time, then to avoid complicacy, every node must inform its monitor that it is in its zone by a periodic control packet named "monitor alive request". This message does not continue in the "hello" session, in fact, when a "hello" session starts it stops and starts when the "hello" session stops with the information given by the "hello" session. The monitor will reply this request with the "monitor alive request repeat" packet only to the node from where it received the request. If the monitor does not receive piggy backed information within a predefined amount of time and no active data delivery is in that session or no monitor alive request packet from a neighbor node in that session then it remove that node from the neighbor node list and if the total number of neighbors is less than the predefined number then it starts a new "hello" session.

In this approach, for the data transfer there are a couple of cases to be considered.

1. If the destination (D) is directly connected with the source (S) (Fig. 1) then the data is simply sent to destination

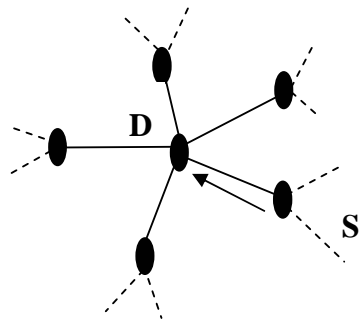

Fig. 1: Source is directly connected to destination.

2. If the destination is not directly connected then the following cases may appear:

(a) If the source (S) finds that the destination is not directly connected to it and the route of the destination is not in the cache then it sends the data to its monitor (M) and its monitor on behalf of it finds the desired destination. If the destination is directly connected to the monitor then it sends the data on behalf of the source (Fig. 2).

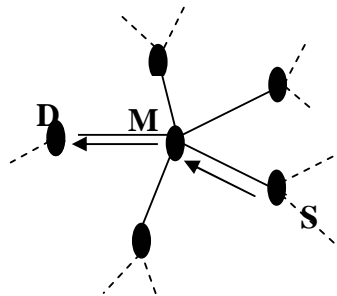

Fig. 2: Source is not directly connected to destination but monitor does.

(b) Now, if the monitor that is requested by a source node to send data on behalf of that node, finds that the destination is not directly connected to it then it searches the cache to find that route and once found data is send by that route. Here, the monitor getting a request from source node finds that it is not directly connected to the destination. So it searches the cache and getting the route it delivers data towards that way (Fig. 3).

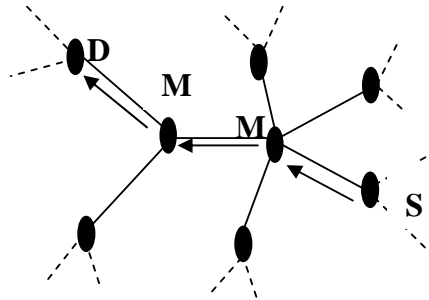

Fig. 3: Monitor has the destination in its cache.

(c) If the source node finds that the route to the destination is in its cache then it follows the route (Fig. 4) but if the route gets broken dynamically, the monitor currently has the packet, propagates queries to all the directly connected monitors of it about the destination (Fig. 5). This process continues upwards. If any of them has it, then they may follow the reverse path to reply the query made by the monitor. If multiple paths are found then the monitor takes the path from which the reply came first, it keeps that path in its cache. In this way protocol recover the dynamically broken link.

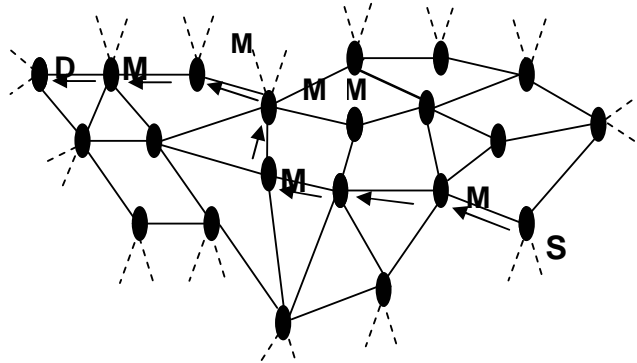

Fig. 4: Source has the destination in its cache.

(d) If the route wanted by the source does not evolve in the route cache of the monitor, it propagates queries to all the directly connected monitors of it about the destination node. This process continues upwards. If any of them has it, then they may follow the reverse path to reply(R) the query made by the monitor. If multiple paths are found then the monitor takes the path from which the reply(R) came first. It keeps the path in its cache. Here as the source request the monitor and the monitor finds that there is no such route in its cache, it then passes the packets to all neighboring monitors and via one of it finds the desired route (Fig. 6).

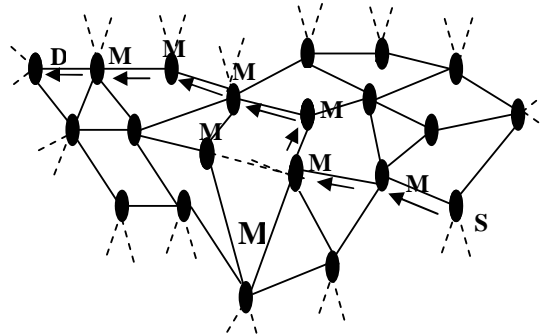

Fig. 5: Dynamically route broken and route discovery.

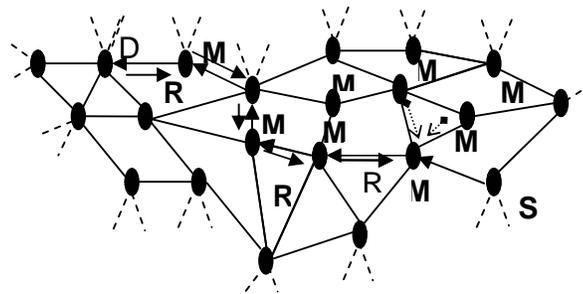

Fig. 6: Route query made by monitor.

(e) If the monitor is querying for destination by asking all the neighboring monitors (M), but the monitor $M_k$ finds that it is in certain range from the source that exceeds the certain limit so that monitor $M_k$ sends a "destination unreachable" (u) message to the source (Fig. 7).

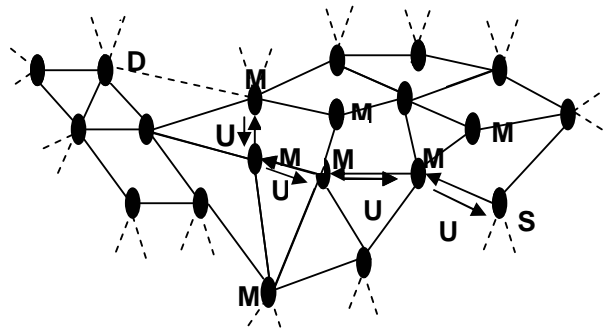

Fig. 7: Monitor forwards "destination unreachable" message.

Existing routing protocols for ad hoc networks can be classified as

- Static vs. Adaptive [16]

- Proactive vs. Reactive [15]
- Hybrid [15]

In static algorithms, the route used by source-destination pairs is fixed regardless of traffic conditions. It can only change in response to a node or link failure. This type of algorithm can't achieve high throughput under abroad variety of traffic input patterns. Most major packet networks uses some form of adaptive routing where the routes used to route between source-destination pairs may change in response to congestion or any other cases where link or path is not found or the link or path is dynamically broken [10].

Proactive or table driven protocols continuously evaluate the routes within the network, so that when a packet needs to be routed is already known and can be immediately used. A node propagates routing information among its neighbors whenever there is a change in its link. This information causes other nodes to re-compute their routing tables. It has routes from a node to every other node in the system. Hypothetically, the topology is a click in the graph theory. So, it is expensive as route construction takes place even though a node does not need it. Proactive protocols are cursed as they waste limited wireless bandwidth of the ad hoc networks. Examples of proactive routing protocols are DSDV, TBRPB, WRP, etc.

Reactive or on-demand protocols invoke a route determination procedure on an on-demand basis by flooding the network with the route query. When a node wants to communicate with another node, it first tries to discover a good route to the destination on which data packets are forwarded. Sending node utilizes a route if it is not damaged or broken. Problem occurs when poor radio signal exists. Whenever a node finds that its link to the next hop is broken, it will send a route error packet back to the source node. This causes waste of available wireless bandwidth as well as a routing delay which results in the increase in latency. The examples of reactive routing protocols for ad hoc networks are AODV, DSR, TORA, etc. Every reactive protocol has the three basic steps:

- Flooding
- Data forwarding and
- Route maintenance.

The on-demand discovery of routes can result on much less traffic than the pro-active schemes, especially when innovative route maintenance schemes are employed. However, the reliance on flooding of the reactive schemes may still lead to a considerable volume of control traffic in the highly versatile ad hoc networking environment. Moreover, because this traffic is concentrated during the periods of route discovery, the route acquisition delay can be significant.

Hybrid protocol refers to the combination of the strengths of several protocols. In most basic hybrid network, one of the protocols would be selected based on its suitability for the specific network's characteristics. Although not an elegant solution, such a framework has the potential to perform as well as the best suited protocol for any scenario, and may outperform either protocol over the entire ad hoc network [12].

From the overview of AMBR protocol, we have found that the AMBR supports suitable features of adaptive and hybrid protocols, So, we can call it adaptive and hybrid and I think that in every cases we will find the best result.

AMBR protocol uses the following symbols:

$S$=Source,  $M$=Monitor,  $D$=Destination
$DL_{max}$ =Maximum Depth Level
$A \rightarrow B$ = A sends data/message to B
$N(U)$ =Neighbor set of node U
$C(U)$ =Set of routes in the cache of node U
$M_{source}$ = Monitor connected to the source node
$NM(M_i)$ =Set of Neighbor Monitors of Monitor $M_i$
$ID(M_k, M)$ =Depth between the Monitor(M) and Monitor($M_k$)

Overall routing technique of the AMBR protocol is described in AMBR algorithm:

### AMBR Algorithm:

```
if D ∈ N(S) then
   S → D       //Directly connected
end if
else if D ∉ N(S) and D ∉ C(S) then
   S → M    // Send data to Monitor
     if D ∈ N(M) then
        M → D //Route through Monitor
     end if
     else if D ∉ N(M) then
        if D ∈ C(M) then
           M → D // Send data through cache
                 //  route of Monitor
        end if
        else if D ∉ C(M) then
           call RouteFinder(M,M,S)
        end else if
     end else if
end else if
else if D ∉ N(S) and D ∈ C(S) then
   while(!RouteBrokenDynamically)
        S → D  //Route in the cache of source
   end while
   if RouteBrokenDynamically then
      call RouteFinder(M,M,S)
   end if
end else if
```

**Function** RouteFinder($M_{source}$,*M*,*Source*)

[This user defined function finds route between the $M_{source}$ and the Monitor connected to the destination node]

```
M → NM(M)
for every element M_i ∈ NM(M) do
    if ID(M_i, M_source) > DL_max then
        M_i → DestinatioNUnreachableMessage(S`)
            return(1)   //Destination is unreachable
    end if
    else if D ∈ N(M_i) then
        Query reply to the M_source and store that path in the cache of the M_source
        return(1) //control back to the main algorithm
    end else if
    else
        call RouteFinder(M_source, M_i, Source)
    end else
end for
```

## 4. ANALYTICAL MODEL

We assume that there are $n$ nodes in the system, and all the nodes have the same distribution of moving speed and direction and the same transmission range $r$. We assume that:

1. The average route length between the source and destination is $E_L$
2. The duration of the packet arrival is exponentially distributed with mean $1/\lambda$.
3. The time between location changes for each node is exponentially distributed with mean $1/\mu$.
4. All $n$ mobile hosts in the network have the same transmission range $r$.

Then, the probability of a route is broken [1],

$$P_B = \frac{\mu}{(\mu + \lambda)} \qquad (1)$$

and the probability that a route is not broken is $(1 - P_B)$.

### 4.1 Packet Routing Probabilities

*Theorem 1.* The probability $P_N$ that at least one of the $E_N$ monitors is able to route from source to destination is,

$$P_N = 1 - (1 - (1 - P_B)^3)^{E_N} \qquad (2)$$

*Proof.* Let the packet is sent from node $A$ to $C$, where $C$ is not directly connected with $A$. So, an on demand diagnosis approach invokes a monitor $M$. The probability of $M$ to find the desired route is $\rho = (1 - P_B)^3$. If $E_N$ be the number of monitors in the network, then $K$ be the number of nodes that want a route by monitor $M$ is given by,

$$P(K) = (^{E_N}C_K) \rho^K (1-\rho)^{E_N - K} \tag{3}$$

Thus, the probability that at least one of the $E_N$ monitors is able to route from source to destination is given by the above.

### 4.2 AMBR Probabilities

*Theorem 2*. The probability $P_R$ that a route discovery succeeds in AMBR protocol is,

$$P_R = 1 - (1-P_0)^K (1-P_0)^{KE_N} \tag{4}$$

*Proof.* If $K$ hops are counted in the case of a failure of route discovery without asking the monitors, then the probability that self diagnosis fails is,

$$P_{F_0} = (1-P_0)^K \tag{5}$$

Here, $P_0$ is the probability that next desired node is found. Again, the probability that total number of $E_N$ monitors also fails to discover the route is,

$$P_{F_1} = (1-P_0)^{KE_N} \tag{6}$$

then, the probability that the route recovery succeeds is,

$$P_R = 1 - P_{F_0} P_{F_1} = 1 - (1-P_0)^K (1-P_0)^{KE_N}$$

*Theorem 3*. The probability that a packet is successfully routed by our protocol is,

$$\begin{aligned}P_S &= (1-P_{F_0}^{1/k})^{E_L} + (E_L - k)(1-P_{F_1}^{1/kE_N})^{(E_L-\hat{k})} \\ &+ (E_L - \hat{k})(1-P_{F_1}^{1/kE_N})^{(E_L-\hat{k})} \\ &+ (1-P_B)^k [1-(1-P_0)^k (1-P_0)^{KE_N (E_L-\hat{k})}]\end{aligned} \tag{7}$$

*Proof.* A packet arrives in one attempt if it passes along all links without being resent by the original host again. That means that an error does not occur along the whole route and an error occurs in one link and the recovery mechanisms are launched and give the result. Therefore, we have

$$\begin{aligned}P_S &= (1-P_{F_0}^{1/k})^{E_L} + (E_L - k)(1-P_{F_1}^{1/kE_N})^{(E_L-\hat{k})} \\ &+ (E_L - \hat{k})(1-P_{F_1}^{1/kE_N})^{(E_L-\hat{k})} \\ &+ (1-P_B)^k [1-(1-P_0)^k (1-P_0)^{KE_N (E_L-\hat{k})}]\end{aligned} \tag{8}$$

## 5.  SIMULATIONS

**5.1 Simulation Environment**

In order to analyze the performance of the proposed AMBR protocol, we run the simulation under the NS-2 testbed with a Carnegie Mellon University (CMU) wireless extension. The simulator parameters are listed in Table 1.

Table 1: Simulation Parameters.

| Parameter | Value |
|---|---|
| Simulator | NS-2(Version 2.29) |
| Network Area | 1300 m × 1300 m |
| Transmission Range | 250 m |
| Bandwidth | 5 kbps |
| Data Packet Size | 512 bytes |
| MAC Layer | IEEE 802.11 |

The network area is confined within 1300 m × 1300 m. Each node in the network has a constant transmission range of 250 m. Constant-bit-rate (CBR) traffic sources are used. The source-destination pairs are spread randomly over the network. Only 512 byte data packets are used. The movement pattern of each node follows the random way-point model. Each node moves to a randomly selected destination with a constant speed between 0 and maximum speed $V_{max}$. When it reaches the destination, it stays there for a random period and starts moving to a new destination. Through out the simulation we calculate the total routing overhead per node.

**5.2 Affected Parameters** [11]

We consider the following parameters that affect the performance of a routing protocol:

*5.2.1   Network size (n).*

The number of nodes in a network determines the density of the network. A dense network will cause more collision and contention.

*5.2.2   Mobility of the node ($V_{max}$).*

The mobility of the node affects the performance of the routing operation. The faster the node moves, the higher is the possibility of the node to lose the information of the neighbors and the information of the monitor.

*5.2.3　Pause time (p).*

At the start of the simulation each node waits for a pause time. It then randomly selects its destination and moves towards this destination with a speed randomly lying between 1 to $V_{max}$, where $V_{max}$ is the maximum speed of a node. Once the destination is reached, another random destination is targeted after a pause.

## 5.3 Results and Analysis

*5.3.1 Sensitivity to Network Size.*

Figure 8 is the corresponding graph to the routing overhead versus network size where $V_{max}$ is 10 meters per second, (m/s), pause time is 10 second, Total simulation time is 200 seconds and the data traffic load constant-packet-rate (CPR) is 10 packets per second (pkt/s).

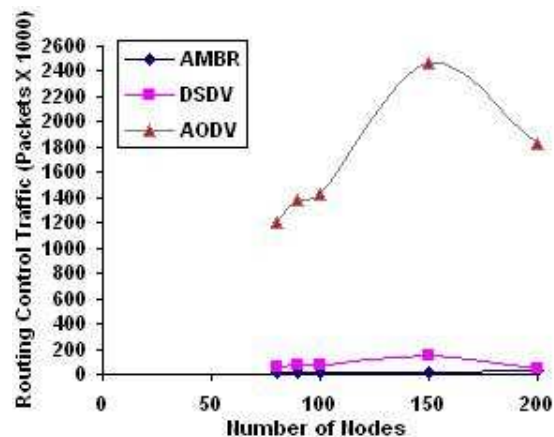

Fig. 8: Routing packet overhead versus size of the network.

We compare the routing overhead against AMBR, DSDV [8] and AODV [9] by varying the network size 80, 90, 100, 150, 200. From the graph it can be easily understood that the routing packet overhead of AODV, DSDV and AMBR increases as network size increases but increasing rate in the case of AODV and DSDV is much higher than the AMBR. Even in denser network (network size 200 nodes) the graph shows that the routing packet overhead of AMBR is much less comparing with the other two protocols. We know that the main aspects of a reactive protocol are flooding, data forwarding and route maintenance. As a result, its (e.g. AODV) routing packet overhead is more. And though proactive protocols continuously evaluate the routes within the network, its (e.g. DSDV) routing packet overhead is as much as AODV. Our protocol is hybrid and adaptive and it has suitable features (mentioned above) compared to others – reactive or proactive. So, its routing packet overhead will be less and here we have found less overhead compared to AODV and DSDV.

*5.3.2 Sensitivity to Mobility of the Node.*

Figure 9 represents the graph showing the effect of the node's mobility on the performance of routing operation. In this case, *n=100, p=10s*, total simulation time=*200s* and *CPR=10* packets per second (pkt/s). This graph shows the routing control packet overhead for different mobility of nodes. Here, we vary the maximum speed of the nodes from 10 m/s to 50 m/s with an interval of 10 m/s.

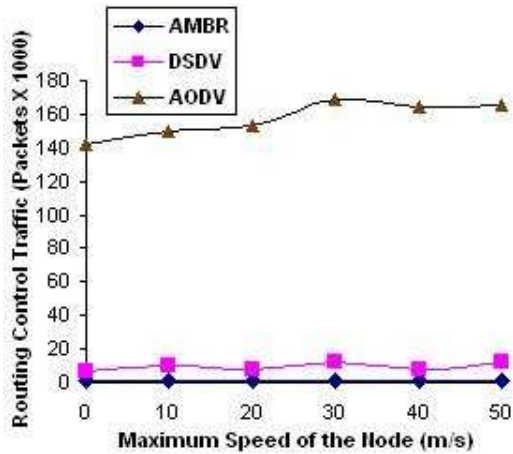

Fig. 9: Routing Packet overhead versus mobility of the node.

From Fig. 9, we see that routing packet overhead is affected by the mobility of the node. When mobility of the node increases, routing control packet overhead is also increase but increasing rate in the case of AMBR is much less than the other two protocols. From this simulation, we conclude that AMBR performs well even when the mobility of the node changes. In case of AODV and DSDV, as the size of the network was increasing, the routing packet overhead was also increasing. And here we have considered mobility of the nodes to observe the performance of AMBR comparing with others such as AODV and DSDV. It is well-known that the mobility of the node affects the performance of the routing operation. The faster the node moves, the higher is the possibility of the node to lose the information of the neighbors and the information of the monitor. In proactive protocol a node propagates routing information among its neighbors whenever there is a change in its link. This information causes other nodes to re-compute their routing tables. It is expensive as route construction takes place even though a node does not need it. Proactive protocols waste limited wireless bandwidth. Reactive protocol also causes waste of available wireless bandwidth as well as a routing delay which results in the increase in latency. So, similar fashion will occur in case of AODV and DSDV. The AMBR protocol improves such features as mentioned in the protocol portrayal section. And that is why AMBR has less overhead than AODV and DSDV.

*5.3.3 Sensitivity to Pause Time.*

In order to investigate the effect of the pause time on the performance of the AMBR protocol, we set the pause time at 50 s, 100 s, 150 s, 200 s, 250 s, 300 s and 350 s. In this case we set, total simulation time=*7500s*, and *CPR=5* packets per second (pkts/s). In comparing the AMBR with DSR-LRR [1], DSR [1] and DSDV we set, *n=50*. In Fig.3 we compare routing control packet overhead for different pause time. In this case we found that AMBR generates lower routing control packet overhead than the other three protocols in different pause time.

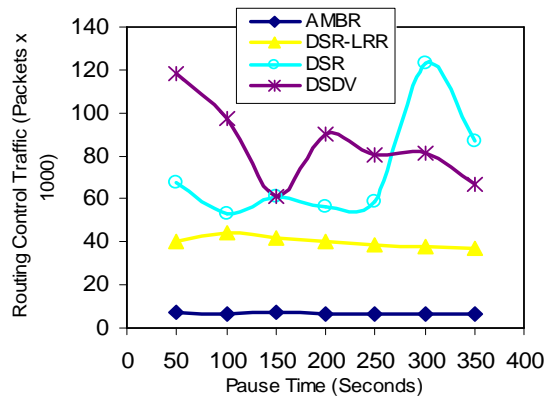

Fig. 10: Routing packet overhead versus pause time.

In the previous two cases we have found that the performance of AODV and DSDV is poor. And now pause time is considered to compare the performance of AMBR protocol with others. When the links in the network break dynamically then AMBR protocol discovers route dynamically from the immediate node linked with the broken link instead of seeking destination node from the source again. Consequently, it will incur less routing overhead in different pause time and its performance will be high. From the figure we have found that as the pause time increases, the performance of other protocols (e.g. DSDV, DSR, DSR-LRR) is degraded compare to AMBR.

## 6. CONCLUSION

In this paper, we performed a through analysis of AMBR protocol. From the analysis we found that AMBR is an efficient technique to route data from source to destination with a low routing cost. Firstly, we proposed an improved analytical model which illustrates the Packet routing probabilities and AMBR probabilities. Secondly, using NS-2 simulator and varying different affecting parameters we measured the routing overhead for AMBR including DSR-LRR, DSR, DSDV and AODV, and compared with each other. These thorough analyses show the inherent strength of the AMBR protocol and firmly determine that AMBR protocol is the most feasible protocol for the ad hoc networks.

For the selection of Monitor, a node should have at least some constant predefined minimum number of neighbors. Future work may introduce a technique from which an optimal number can be found dynamically for different network conditions. To alleviate the network from extravagant periodic routing control traffic, our target was to introduce event driven packets as much as possible but still some messages are partially periodic like 'Monitor Alive Request' packets as they depends on some events. So as a future work a technique can be introduced that can turn these partially periodic packets into fully event driven packets.